\documentclass[aip,jcp,preprint,amsmath,amssymb,floatfix]{revtex4-1}

\usepackage{graphicx}
\usepackage{multirow}
\usepackage{braket}
\usepackage{algorithm}
\usepackage{setspace}
\usepackage{url}
\usepackage[monochrome]{color}

\newcommand{\bs}[1]{\boldsymbol{#1}}
\newcommand{\brat}[1]{\bra{\tilde{#1}}}

\newcommand{\ta}[0]{\tilde{a}}
\newcommand{\ad}[0]{a^\dag}
\newcommand{\bkappa}[0]{\boldsymbol{\kappa}}
\newcommand{\der}[2]{\frac{\partial{#1}}{\partial{#2}}}
\newcommand{\ku}[0]{\kappa^u}
\newcommand{\kd}[0]{\kappa^d}
\newcommand{\tamu}[0]{\tau_{\mu_n}}
\newcommand{\lamu}[0]{\lambda_{\mu_n}}
\newcommand{\tH}[0]{\tilde{H}}
\newcommand{\half}[0]{\frac{1}{2}}

\newcommand{\Eval}[1]{\left.#1\right\rvert_{\bkappa = \boldsymbol{0}}}
\newcommand{\de}[1]{\delta_{#1}}

\newcommand{\La}[0]{\Lambda}
\newcommand{\la}[0]{\lambda}
\newcommand{\al}[0]{\alpha}
\newcommand{\be}[0]{\beta}
\newcommand{\ti}[1]{\tilde{#1}}
\newcommand{\tPsi}[0]{\tilde{\Psi}}

\newcommand{\Hzz}[0]{H_{00}}
\newcommand{\Htt}[0]{H_{22}}
\newcommand{\Htz}[0]{H_{20}}
\newcommand{\Hzt}[0]{H_{02}}

\begin{document}

\title{Demonstrating that the nonorthogonal orbital optimized coupled cluster model converges to full 
configuration interaction} 
\author{Rolf H. Myhre}
\email[email:]{r.h.myhre@kjemi.uio.no}
\affiliation{Hylleraas Centre for Quantum Molecular Sciences, Department of Chemistry, 
University of Oslo, 0315 Oslo, Norway}

\date{\today}

\begin{abstract}

Coupled cluster (CC) methods are among the most accurate methods in quantum chemistry. 
{\color{red}However, the 
standard CC linear response formulation is not gauge invariant resulting in errors when modelling 
properties like optical rotation and electron circular dichroism. Including an explicit unitary orbital 
rotation in the CC Lagrangian makes the linear response function gauge invariant, but the resulting 
models are not equivalent to full configuration interaction (FCI) in the untruncated limit. In this 
contribution, such methods are briefly discussed and it is demonstrated that methods using a 
nonorthogonal orbital transformation, such as nonorthogonal orbital optimized CC (NOCC), can converge to 
FCI in the untruncated limit. This has been disputed in the literature.}

\end{abstract}

\maketitle

\section{Introduction}
{\color{red}Coupled cluster (CC) theory is the most accurate tool in regular use for describing 
molecular systems with an electronic wavefunction dominated by a single reference 
determinant\cite{CCreview}. Such systems include most molecules in their ground state minimum energy 
geometry. The method can also be used to describe electronically excited states using the linear 
response\cite{lin_resp_fun,CCRespFunc} (LR) or closely related equation of motion\cite{EOMCC} (EOM) 
formalism. Due to the high accuracy of the model, current research is focused on reducing its relatively 
high computational cost and expanding it to systems with multireference character. Efforts in the former 
has focused on exploiting the short range of electron correlation to reduce the scaling of the CC models 
and has been quite successful\cite{PAO_rev,CrawfordLocalCC,LPNOseries3,DECrev}. Many models has also 
been proposed to solve the multireference problem, but they tend to suffer from very high computational 
cost, instabilities or low accuracy\cite{StateSpecCC,MultiRev}. 

Another issue that has received less attention in the literature is the fact that standard 
truncated CC is not gauge invariant, even in the complete basis limit\cite{CCRespRev}. This is 
a consequence of truncated CC not satisfying the conditions of the Ehrenfest theorem, eq. 
\eqref{ehrenfest}, and results in discrepancies in properties like transition moments when using 
different gauges, for example dipole length and dipole velocity. 
\begin{equation}\label{ehrenfest}
   \frac{d}{dt}\braket{A} = i\braket{[H,A]} + \left<\der{A}{t}\right>
\end{equation}
Typically, the discrepancies are quite small if the method provides a reasonable good 
description of the wavefunction and the basis set is sufficiently large\cite{GaugeCC}. However, for 
properties depending on magnetic fields such as optical rotation (OR) and electron circular 
dichroism, translation in space is a gauge transformation. Consequently, the results of such 
calculations will depend on the placement of the origin in the dipole length gauge which is 
completely unphysical\cite{OptRotCrawford1,OptRotCrawford2}. In the dipole velocity gauge, CC OR 
calculations are origin independent, but includes an unphysical zero-frequency contribution. 
In the modified velocity gauge, the zero-frequency contribution is subtracted resulting in consistent 
results at the cost of an additional calculation\cite{MVG}. Note that basis set incompleteness is also 
a cause of gauge dependence, but this can be avoided using gauge including atomic 
orbitals\cite{LondonOrbs,GIAO}.

While standard CC theory does not satisfy the conditions of the Ehrenfest 
theorem\cite{CCRespRev}, this can be achieved for one-electron operators by including an explicit 
orbital transformation in the CC Lagrangian\cite{GaugeInvariant,NOCC}. In standard CC, the right hand 
singles cluster operator, $T_1$ acts as an approximate orbital transformation\cite{CCOptOrb87} and is 
removed when including an explicit orbital transformation because it is redundant. However, it is less 
clear what to do with the left hand singles cluster operator $\La_1$ and several methods have been 
proposed. Note that for single reference systems, the orbital transformations are small and only results 
in small changes in total energy. However, for some systems where multireference character is caused by 
orbital instabilities, significant improvements can be observed\cite{MultiRev,OCC}.
}

Brueckner CC\cite{Brueckner,BCC,BCCresp} (BCC) was, like standard CC\cite{nuclear_cc1}, originally 
developed in nuclear physics. Orbital rotation parameters are included in the BCC Lagrangian 
exponentially, ensuring unitary transformations and orthogonal orbitals. The solution to the BCC 
equations is a wavefunction that is invariant with respect to the cluster amplitudes and the orbitals 
are rotated to a basis where the singles right hand cluster operator is zero. The resulting model 
satisfies the Ehrenfest theorem, but unphysical second order poles appear in the response function. This 
and the fact that only small improvements are observed for the ground state energy compared to standard 
CC has limited the application of this model.

Orbital optimized coupled cluster\cite{OCC,QCOCC,GaugeInvariant} (OCC) is similar to BCC and is also 
referred to as variational Brueckner CC. In this method, both the left and right hand singles amplitudes 
are set to zero and the orbital parameters are obtained by minimizing the energy. In this way, one 
obtains a response function with the correct pole structure. However, as pointed out by K{\"o}hn and 
Olsen\cite{OCCfail}, this method is not equivalent to full correlation interaction (FCI) in the 
untruncated limit. {\color{red} Nonetheless, the advantages of OCC has led to continued development of 
CC models with orbital optimization. For example, Crawford \emph{et al.} proposed a CC model that 
combined orbital optimization with orbital localization in order to reduce the scaling of CC 
calculations for OR\cite{LocalOptCC}. In perfect paired CC, only cluster amplitudes involving paired 
electrons are retained\cite{PPCC}. This greatly reduces the number of amplitudes and computational cost. 
Scuseria \emph{et al.} combined the formalism with orbital optimization and demonstrated that the method 
gave good results, even for strongly correlated systems were standard CC fails\cite{OPPCC1,OPPCC2}. 
Recently, Head-Gordon \emph{et al.} expanded the model to include paired quadruples and even hextuples 
in an active space\cite{OPPCCQ1,OPPCCQ2}. Note that the accuracy was improved when including singles in 
the cluster operator.}

The nonorthogonal orbital optimized CC (NOCC) approach is similar to OCC in that both sets of singles 
amplitudes are redundant and set to zero\cite{NOCC}. As implied by the name, the difference lies in 
the orbital transformation. By relaxing the demand for a unitary transformation, the orbitals are no 
longer orthogonal, but biorthogonal, resulting in a bivariational Lagrangian\cite{varCC}. In their 
paper, K{\"o}hn and Olsen conjectured that NOCC would suffer the same defects as OCC and not reach 
the FCI limit due to the lack of singles amplitudes. In this contribution, we will demonstrate that this 
is not the case and untruncated NOCC is equivalent to FCI.

Two other methods are worth a brief mention before we proceed. In the extended CC 
(ECC)\cite{Arponen,ECCseries1} method by Arponen, both the excitations and deexcitations are 
parametrized exponentially. {\color{red}Arponen showed that the standard CC model can be viewed as an 
approximation to ECC where the exponential of the deexcitations has been replaced by a linear 
parametrization that becomes identical in the FCI limit.} The ECC Lagrangian is fully bivariational, 
ensuring the uniqueness and existence of a solution as well as bounds for the error\cite{ecc_err}. 
Unfortunately, this formulation results in an enormous number of terms, making a working implementation 
unfeasible. Another method based on the bivariational approach is the orbital adapted CC (OACC) method 
by Kvaal\cite{OACC}. This method is similar to NOCC, but the left and right hand side orbitals are 
allowed to span different subspaces of the total orbital space, further increasing flexibility.

\section{NOCC equations}

Nonorthogonal OCC is differentiated from standard OCC by the use of a biorthogonal instead of 
orthogonal basis and we will start our discussion with the rotational orbital parametrization. 
{\color{red}All expressions are in the spinorbital basis and we only consider the untruncated FCI limit.}
In order to ensure a unitary transformation, an exponential parametrization is employed in OCC. 
The orthogonal reference creation, $\hat{a}^\dag_p$, and annihilation, $\hat{a}_p$, operators and 
reference state function $\ket{\hat{\phi}}$ are transformed according to eq. \eqref{unitrans}.
\begin{equation}\label{unitrans}
   \begin{aligned}
      a^\dag_p &= \exp(-\bkappa)\hat{a}^\dag_p\exp(\bkappa)\\
      a_p &= \exp(-\bkappa)\hat{a}_p\exp(\bkappa)\\
      \ket{\phi} &= \exp(-\bkappa)\ket{\hat{\phi}}
   \end{aligned}
\end{equation}
By demanding that $\bkappa$ is antihermitian, the resulting transformation is unitary.
\begin{equation}\label{unikappa}
   \bkappa = \sum_{pq}\kappa_{pq}a^\dag_{p}a_{q}, \quad \bkappa = -\bkappa^\dag
\end{equation}
It can be shown that rotations between two occupied or two virtual orbitals in the reference 
wavefunction are redundant in OCC, so only the off-diagonal blocks corresponding to occupied-virtual 
and virtual-occupied rotations are included in $\bkappa$. 

In NOCC, the requirement that $\bkappa$ is antihermitian is removed, resulting in a non-unitary 
transformation of the orbitals. Equation \eqref{unitrans} is still valid, but the creation and 
annihilation operators are no longer each other's complex conjugates. To emphasize this, we will 
label the annihilation operator and left hand side with a tilde.
\begin{equation}\label{nunitrans}
   \begin{aligned}
      (\tilde{a}_p)^\dag &= \big(\exp(-\bkappa)\hat{a}_p\exp(\bkappa)\big)^\dag\\ 
                         &= \exp(\bkappa^\dag)\hat{a}^\dag_p\exp(-\bkappa^\dag)\\
                         &\neq\exp(-\bkappa)\hat{a}^ \dag_p\exp(\bkappa) = a^\dag_p
   \end{aligned}
\end{equation}
Despite not being conjugates of each other, the anticommutation relations holds and we can employ 
Wick's theorem in the standard way\cite{LowdinQtheory2,NUBogoliubov}. Occupied-occupied and 
virtual-virtual orbital rotations are also still redundant and we label the excitation and deexcitation 
parameters in $\bkappa$ with $u$ for up and $d$ for down for convenience. We use the standard notation 
where indices $i,j,k,\ldots$ and $a,b,c,\ldots$ refer to occupied and virtual orbitals in the reference 
state respectively.
\begin{equation}\label{nunikappa}
   \bkappa = \sum_{ai}\kappa^u_{ai}a^{\dag}_{a}\ta_{i} + \kappa^{d}_{ia}a^{\dag}_{i}\ta_{a}
           = \sum_{ai}\kappa^u_{ai}X_{ai} + \kappa^{d}_{ia}Y_{ia}
\end{equation}

{\color{red}In eq. \eqref{nunikappa} we have introduced the right hand, $X_{ai}$, and left hand, 
$Y_{ia}$, singles excitation operators. Higher excitation operators are similarly defined, analogously 
to standard CC theory. In an orthogonal basis these are each other's complex conjugates, but this is 
not generally true in a biorthogonal basis, $Y_\mu \neq X^\dag_\mu$, and we need different symbols for 
the operators.}

The starting point for the NOCC model is the bivariational NOCC Lagrangian $L$. {\color{red}
\begin{equation}\label{NOCC_Lag}
\begin{gathered}
   L = \brat{\Psi}H\ket{\Psi} \\
     = \brat{\phi}(1 + \Lambda)\exp(-T)\exp(-\bkappa)H\exp(\bkappa)\exp(T)\ket{\phi}
\end{gathered}
\end{equation}
Explicitly including the orbital transformation terms in the derivation of the NOCC equations would 
result in extremely complicated expressions because $\bkappa$ does not commute with $T$ or $\Lambda$.
We therefore express the equations in the optimized basis where $\bkappa = \boldsymbol{0}$ and a 
solution to the Schr{\"o}dinger equation corresponds to a stationary point of the Lagrangian. From 
an implementation perspective, this can be viewed as expanding the exponentials of $\bkappa$ and only 
keeping zero order terms. This suggest an algorithm that iterates between orbital transformation and 
amplitudes until self consistency\cite{QCOCC}.}

{\color{red}The exponential parametrization of the orbital rotations ensures that our basis and manifold 
of states are biorthogonal and we assume unit overlap between the reference bra and ket states.
\begin{equation}\label{biortozero}
   \braket{\tilde{\mu}|\phi} = \braket{\tilde{\phi}|Y_{\mu}|\phi} = 0 \quad
   \braket{\tilde{\phi}|\mu} = \braket{\tilde{\phi}|X_{\mu}|\phi} = 0
\end{equation}
\begin{equation}\label{biortoone}
   \braket{\tilde{\phi}|\phi} = 1 \quad
   \braket{\tilde{\mu}|\nu} = \delta_{\mu,\nu}
\end{equation}
$T$ and $\Lambda$ are defined similarly to the standard cluster operators except the redundant singles 
excitations have been omitted.}
\begin{equation}\label{TLop}
   T = \sum_{\mu_n}\tau_{\mu_n}X_{\mu_n} \quad \Lambda = \sum_{\mu_n}\lambda_{\mu_n}Y_{\mu_n} 
   \quad n \geq 2
\end{equation}
In eq. \eqref{TLop}, $\tau_{\mu}$ and $\lambda_{\mu}$ are the amplitude parameters of the operators.  
{\color{red}The NOCC left, $\brat{\Psi}$, and right, $\ket{\Psi}$, 
wavefunctions must satisfy the standard CC equations, including the singles projection, in the 
biorthogonal basis in order to be eigenfunctions of the Hamiltonian. Note that the Hamiltonian is not 
Hermitian in this basis.
\begin{equation}\label{rprojection}
\begin{gathered}
   \brat{\mu}\exp(-T)H\ket{\Psi} = E\brat{\mu}\exp(-T)\ket{\Psi}\\ 
   = E\brat{\mu}\exp(-T)\exp(T)\ket{\phi} = E\braket{\tilde{\mu}|\phi} = 0
\end{gathered}
\end{equation}
\begin{equation}\label{lprojection}
\begin{gathered}
   \brat{\Psi}[H,Y_{\mu}]\ket{\Psi} = \brat{\Psi}HY_{\mu}\ket{\Psi} - \brat{\Psi}Y_{\mu}H\ket{\Psi}\\ 
 = E(\brat{\Psi}Y_{\mu}\ket{\Psi} - \brat{\Psi}Y_{\mu}\ket{\Psi}) = 0
\end{gathered}
\end{equation}}
 
At the stationary point, the differential of $L$ must be zero with respect to the four sets of 
parameters: $\{\tau\}, \{\lambda\}, \{\ku\}$ and $\{\kd\}$ resulting in four sets of equations.
\begin{align}
   \der{L}{\lamu} &= \bra{\tilde{\mu}_n}\exp(-T)H\exp(T)\ket{\phi}\label{lder}\\
   \der{L}{\tamu} &= \brat{\phi}(1+\Lambda)\exp(-T)[H,X_{\mu_n}]\exp(T)\ket{\phi}\label{tder}\\
   \der{L}{\ku_{\mu_1}} &= 
   \brat{\phi}(1+\Lambda)\exp(-T)[H,X_{\mu_1}]\exp(T)\ket{\phi}\label{uder}\\
   \der{L}{\kd_{\mu_1}} &= 
   \brat{\phi}(1+\Lambda)\exp(-T)[H,Y_{\mu_1}]\exp(T)\ket{\phi}\label{dder}
\end{align}
In order to prove that NOCC is equivalent to FCI, {\color{red} we must first demonstrate that the above 
equations results in a wavefunction that satisfies all the projection equations. In particular, the 
singles projection equation must be satisfied\cite{OCCfail}.} Furthermore, for the equivalence to go 
both ways, we must prove that the standard CC wavefunction can be rotated to a basis were it satisfies 
the NOCC equations. 

Equations \eqref{lder} and \eqref{tder} are the standard projection equations from CC theory and are 
required in order to satisfy the FCI equation. Similarly, eq. \eqref{uder} is identical to the 
derivative with respect to the right hand singles amplitudes in standard CC. Only eq. \eqref{dder} 
differs from the equivalent projection equations in standard CC and requires further analysis. To 
simplify the analysis, we will introduce some additional notation.
\begin{align}
   \brat{\Lambda} &= \brat{\phi}(1 + \Lambda)\label{lamstat}\\
   \tH &= \exp(-T)H\exp(T)\label{tildeH}
\end{align}
Using the Baker-Campbell-Hausdorff expansion and commuting out the $Y_{\mu_1}$-operator, we are left 
with three terms. 
\begin{equation}\label{threeterm}
   \begin{aligned}
      \der{L}{\kd_{\mu_1}} &= \brat{\Lambda}[\tH,Y_{\mu_1}]\ket{\phi}\\
                     &+ \brat{\Lambda}[\tH,[Y_{\mu_1},T]]\ket{\phi}\\
                     &+ \half\brat{\Lambda}[\tH,[[Y_{\mu_1},T],T]]\ket{\phi}\\
   \end{aligned}
\end{equation}
We analyze the expression in eq. \eqref{threeterm} term by term and start with the first. Writing out 
the commutator we get the standard singles projection with some additional terms.
\begin{equation}\label{firstterm}
\begin{aligned}
    &\brat{\Lambda}[\tH,Y_{\mu_1}]\ket{\phi} =\\
   &-\bra{\tilde{\mu}_1}\tH\ket{\phi} - \sum_n\lamu\bra{\tilde{\mu}_n}Y_{\mu_1}\tH\ket{\phi}
\end{aligned}
\end{equation}
{\color{red} In the FCI limit, any projection against doubles and higher are zero due to eq. 
\eqref{lder}} and eq. \eqref{firstterm} reduces to the standard singles projection equation.
Similarly, the double commutator in the last term reduces to a modified 
cluster operator {\color{red}of triple excitations and higher}.
\begin{equation}\label{thirdterm}
   \brat{\Lambda}[\tH,[[Y_{\mu_1},T],T]]\ket{\phi} =
   \sum_{\mu_n}B_{\mu_n}\brat{\Lambda}[\tH,X_{\mu_n}]\ket{\phi}
\end{equation}
Above, the coefficients $B_{\mu_n}$ are all linear combinations of products of two cluster amplitudes 
and all these terms are zero due to eq. \eqref{tder}.

{\color{red}The second term in eq. \eqref{threeterm} results in three types of terms.
\begin{equation*}\label{singlecommute}
   \begin{aligned}
      &\pm\brat{\Lambda}[\tH,X_{\mu_{n-1}}]\ket{\phi}\\
      &+ \brat{\Lambda}[\tH,X_{\mu_{n-1}}a^\dag_b\ta_a]\ket{\phi}\\
      &+ \brat{\Lambda}[\tH,X_{\mu_{n-1}}\ta_j a^\dag_i]\ket{\phi}
   \end{aligned}
\end{equation*}}
The sign of the first term depends on the order of the creation and annihilation operators, but this 
term is zero anyway due to eqs. \eqref{tder} and \eqref{uder} and we only need to worry about the 
last two terms. When acting on the reference state, the extra creation and annihilation operators become 
zero so we only need the terms with the operator to the left of the Hamiltonian. When including the 
cluster operator with amplitudes, the single commutator term in eq. \eqref{threeterm} takes the form in 
eqs. \eqref{secondterm1} and \eqref{secondterm2}. The compound index $\mu_{n+1}$ differs 
from $\mu_n$ in that it includes an extra excitation involving one external index.
\begin{equation}\label{secondterm1}
   \sum_{\nu_m,\mu_n,b}\lambda_{\nu_m}\tau_{\mu_{n+1}}
   \brat{\phi}Y_{\nu_m}X_{\mu_n}\ad_b\ta_a\tH\ket{\phi}
\end{equation}{\color{red}
\begin{equation}\label{secondterm2}
   \sum_{\nu_m,\mu_n,j}\lambda_{\nu_m}\tau_{\mu_{n+1}}
   \brat{\phi}Y_{\nu_m}X_{\mu_n}\ta_j a^\dag_i\tH\ket{\phi}
\end{equation}}

Equations \eqref{secondterm1} and \eqref{secondterm2} give different results depending on the excitation 
level of $\nu_m$ and $\mu_n$ and there are three different cases: $m \leq n$, $m = n+1$ and $m > n+1$. 
In the first case, the term is zero due to projection and the last case is zero due to eq. \eqref{lder}. 
When $m = n+1$, the term becomes a linear combination of the single projections with one index differing 
from the original external indexes. 

\begin{equation}\label{nonzero}
\begin{gathered}
   \brat{\Lambda}[\tH,[a^\dag_i\ta_a,T]]\ket{\phi} = \\ 
   -\sum_{j}C^{ai}_j\brat{^a_j}\tH\ket{\phi}
   -\sum_{b}C^{ai}_b\brat{^b_i}\tH\ket{\phi} 
\end{gathered}
\end{equation}
The coefficients $C$ are products of the $\lambda$ amplitudes and antisymmetrized cluster amplitudes 
$\tau^{AS}_{\mu_n,i}$.  
\begin{equation}\label{Cbi}
   C^{ai}_j = \sum_{\mu_{n,j}}\lambda_{\mu_{n,j}}\tau^{AS}_{\mu_{n,i}} \quad
   C^{ai}_b = \sum_{\mu_{n,b}}\lambda_{\mu_{n,b}}\tau^{AS}_{\mu_{n,a}}
\end{equation}
Compound indexes of the type $\mu_{n,p}$ indicates that the excited state involves the orbital $p$ 
and the indices $\mu_{n,p}$ and $\mu_{n,q}$ differ only in this index. For example, the doubles 
contributions are sums over three indices.
\begin{equation}\label{example1}
   C^{ai}_j \leftarrow \sum_{bck}\lambda^{bc}_{jk}(\tau^{bc}_{ik} - \tau^{cb}_{ik}) 
             = 2\sum_{bck}\lambda^{bc}_{jk}\tau^{bc}_{ik}
\end{equation}{\color{red}
\begin{equation}\label{example2}
   C^{ai}_b \leftarrow \sum_{cjk}\lambda^{bc}_{jk}(\tau^{ac}_{jk} - \tau^{ac}_{kj}) 
            = 2\sum_{cjk}\lambda^{bc}_{jk}\tau^{ac}_{jk}
\end{equation}}

Adding the terms together, eq. \eqref{dder} reduces to the standard single projection and sums of 
single projections that differ in one index.
\begin{equation}\label{finaldder}
\begin{aligned}
   \Eval{\der{L}{\kd_{ia}}} &= -\brat{^a_i}\tH\ket{\phi}\\
   &- \sum_{j}C^{ai}_j\brat{^a_j}\tH\ket{\phi}
    - \sum_{b}C^{ai}_b\brat{^b_i}\tH\ket{\phi} 
\end{aligned}
\end{equation}
Equation \eqref{finaldder} can also be written on matrix form.
\begin{equation}\label{ddermat}
   \boldsymbol{0} = \boldsymbol{A}\boldsymbol{x}
\end{equation}
\begin{equation}\label{xai}
   x_{ai} = \brat{^a_i}\tH\ket{\phi}
\end{equation}
\begin{equation}\label{Aai}
   A_{ai,bj} = \de{ai,bj} + \de{i,j}C^{ai}_b + \de{a,b}C^{ai}_j
\end{equation}
{\color{red}The structure of $\bs{A}$ is visualized in eq. \eqref{Astruct} where $\times$ indicates a nonzero 
element of the matrices.
\begin{widetext}
\begin{equation}\label{Astruct}
   \renewcommand{\arraystretch}{1.0}
   \bs{A} = \bs{I} + 
   \scalebox{0.6}{$
   \left(
   \begin{array}{*{12}c}
      \times &        &        & \times &        &        & \times &        &        \\
             & \times &        &        & \times &        &        & \times &        \\
             &        & \times &        &        & \times &        &        & \times \\
      \times &        &        & \times &        &        & \times &        &        \\
             & \times &        &        & \times &        &        & \times &        \\
             &        & \times &        &        & \times &        &        & \times \\
      \times &        &        & \times &        &        & \times &        &        \\
             & \times &        &        & \times &        &        & \times &        \\
             &        & \times &        &        & \times &        &        & \times \\
   \end{array}
   \right)
   $} +
   \scalebox{0.6}{$
   \left(
   \begin{array}{*{12}c}
      \times & \times & \times &        &        &        &        &        &        \\
      \times & \times & \times &        &        &        &        &        &        \\
      \times & \times & \times &        &        &        &        &        &        \\
             &        &        & \times & \times & \times &        &        &        \\
             &        &        & \times & \times & \times &        &        &        \\
             &        &        & \times & \times & \times &        &        &        \\
             &        &        &        &        &        & \times & \times & \times \\
             &        &        &        &        &        & \times & \times & \times \\
             &        &        &        &        &        & \times & \times & \times \\
   \end{array}
   \right)
   $}
\end{equation}
\end{widetext}

Clearly, a solution which satisfies all the singles projections, i.e. $\bs{x}=\bs{0}$, will 
satisfy eq. \eqref{ddermat} so $\ket{\Psi}$ will satisfy the NOCC equations if it is an eigenfunction of 
$H$. However, we also have to show that it is unique. In order for a matrix equation like eq. 
\eqref{ddermat} to have a unique solution, $\bs{A}$ must be nonsingular with $\det(\bs{A}) \neq 0$. If 
the FCI wavefunction is dominated by a single determinant, the amplitudes in $T$ and $\Lambda$ will be 
small. Consequently the off-diagonal elements in $\bs{A}$ are much smaller than 1 and $\bs{A}$ will be 
strictly diagonally dominant. Such matrices are never singular and the solution to the NOCC equations is 
unique.

In the multireference case, we can no longer assume that the amplitudes are small. However, we first 
note that the space of singular $\bs{A}$-matrices is one dimensional because such matrices must satisfy 
$\det(\bs{A}) = 0$ and a minuscule change in any coefficient would make it nonsingular. Consequently, 
the chance of generating a singular matrix by choosing the coefficients at random is zero, given 
infinite numerical accuracy. In our case, the coefficients are not chosen at random, but fixed by the 
eqs. \eqref{lder} and \eqref{tder}. However it seems highly unlikely that one would obtain a self 
consistent solution resulting in a singular matrix. We note that the $\Lambda$-amplitudes are 
proportional to the complex conjugate $T$-amplitudes to first order in standard CC with a Hermitian 
Hamiltonian, $\bs{\lambda} \sim \bs{\tau}^\dag$. This is a consequence of the Hamiltonian being 
Hermitian. If the basis transformation remains close to unitary, we can expect the largest coefficients 
to appear on the diagonal in $\bs{A}$ and be positive because these coefficients are the products of 
matching indices in eq. \eqref{Cbi}. In Appendix \ref{minex}, we explore the minimal example of two 
electrons in two orbitals.
}

To complete the proof, we must also show that a standard CC wavefunction rotated to a basis where 
$\lambda_1$ and $\tau_1$ are zero would satisfy the equations. {\color{red}We will now investigate the 
existence and uniqueness of such a rotation using the concept of strong 
monotonicity\cite{Zarantonello,NonLinMonOp,CC_uni1}.} Note that setting $\bkappa = \Lambda_1 - T_1$ will 
remove $\Lambda_1$ and $T_1$ from the cluster operators to first order {\color{red} in $\bkappa$ and the 
amplitudes.} To simplify, we assume our starting basis is 
one where $T_1$ is zero which can always be reached by setting $\bkappa = -T_1$. A function $f$ is said 
to be locally strongly monotone if the function $\Delta(\kappa_1,\kappa_2)$ satisfies eq. \eqref{delta} 
for all $\bkappa_1$ and $\bkappa_2$ on an open set, $b$.
\begin{equation}\label{delta}
\begin{gathered}
   \Delta(\kappa_1,\kappa_2) = 
   \braket{f(\bkappa_1) - f(\bkappa_2),\bkappa_1 - \bkappa_2}\\ 
                             \geq c||\bkappa_1 - \bkappa_2||^2
\end{gathered}
\end{equation}
In equation \eqref{delta}, $\braket{\cdot,\cdot\cdot}$ indicates an inner product {\color{red} and in 
this case it is simply the vector product of the vector function $f$ over the compound indices $ai$ 
and $ia$}. {\color{red}By Zarantonello's theorem, the equation $f(x) = a$ has a locally unique solution on 
$b$ if $f$ is strongly monotone\cite{Zarantonello}.} The vector function $f$ is the same size 
as $\bkappa$ and can be divided in two parts, $f_{ai}$ and $f_{ia}$ that are the projections of $T_1$ and 
$\Lambda_1$ respectively.{\color{red}
\begin{align}
   f_{ai} &= \brat{\phi}Y_{ia}\exp(\bkappa)\ket{\Psi}\label{fai}\\
   f_{ia} &= -\bra{\tPsi}\exp(-\bkappa)X_{ai}\exp(\bkappa)\ket{\Psi}\label{fia}
\end{align}

By expanding $f_{ai}$ and $f_{ia}$ to first order in $\bkappa$ we can write $\Delta$ on 
quadratic form. If the corresponding matrix is positive definite, eq. \eqref{delta} holds.
\begin{widetext}
\begin{equation}\label{deltaM}
   \Delta \approx 
   \begin{pmatrix}
      \Delta\kappa_{ai} & \Delta\kappa_{ia}\\
   \end{pmatrix}
   \begin{pmatrix}
      \boldsymbol{I} & \brat{\phi}\ad_{i}\ta_{a}\ad_{j}\ta_{b}\ket{\Psi}\\
      \bs{0} & \boldsymbol{B}
   \end{pmatrix}
   \begin{pmatrix}
      \Delta\kappa_{bj}\\
      \Delta\kappa_{jb}
   \end{pmatrix}
\end{equation}
\begin{equation}\label{Amat}
   \boldsymbol{B} = \bs{I} - \brat{\Psi}\de{ij}\ad_{a}\ta_{b} + \de{ab}\ta_{i}\ad_{j}\ket{\Psi}
\end{equation}
\end{widetext}}
The upper left block in eq. \eqref{deltaM} is simply the identity matrix while {\color{red} the upper 
right block reduces to $\tau^{ab}_{ij}$. The lower left block is zero, and $\bkappa^u$ leaves $\la_1$ 
unchanged to first order. Finally, the lower right block is the identity plus some additional terms that 
are at least second order in the amplitudes.} Assuming that the reference state dominates the 
wavefunction, the matrix will be positive definite {\color{red} and $\Delta$ is strongly monotone. 
Consequently, there must be a unique orbital rotation of the untruncated CC solution for single 
reference cases that removes $T_1$ and $\Lambda_1$.} Because this wavefunction will satisfy the singles 
projection by definition, eq. \eqref{dder} will also be satisfied and the wavefunction is a solution of 
the NOCC equations. 

In multireference systems, the picture is more complicated. Standard CC explicitly satisfies all the 
projection equations and has the correct solution. However, it is less clear whether it is 
still possible to rotate away $T_1$ and $\Lambda_1$. The form of $\Delta$ in eq. \eqref{deltaM} is 
obtained by approximating the exponential of $\bkappa$ with a linear expansion. This will no longer 
be valid when $T_1$ and $\La_1$ become large and $\Delta$ will become a complicated function depending 
on higher order terms in $\bkappa$. 

It is worthwhile to briefly compare NOCC and OCC. First, we note that the orbital rotation that removes 
both $\Lambda_1$ and $T_1$ from the standard CC wavefunction is not unitary and the untruncated OCC 
wavefunction cannot be a solution to the CC equations in general. Comparing eqs. \eqref{uder} and 
\eqref{dder} with the OCC equivalent we see that enforcing a unitary transformation halves the number of 
transformation parameters and the equivalent of eqs. \eqref{uder} and \eqref{dder} is a single equation.
\begin{equation}\label{OCCder}
   \begin{aligned}
      \der{L}{\kappa_{\mu_1}} &= 
      \brat{\Lambda}\exp(-T)[H,X_{\mu_1} - Y_{\mu_1}]\exp(T)\ket{\phi}\\
      &\Rightarrow  \brat{\Lambda}\exp(-T)[H,X_{\mu_1}]\exp(T)\ket{\phi} \\ 
      &= \brat{\Lambda}\exp(-T)[H,Y_{\mu_1}]\exp(T)\ket{\phi}
   \end{aligned}
\end{equation}
{\color{red}This is sufficient to satisfy the conditions of the Ehrenfest theorem, but does not require 
the terms to be zero on their own.\cite{GaugeInvariant} Consequently, the singles projection 
equations are not satisfied as noted by K{\"o}hn and Olsen\cite{OCCfail}.} For NOCC, the two terms are 
zero independently.

\section{Conclusion}

{\color{red}
In this contribution we have demonstrated that NOCC is equivalent to FCI in the untruncated limit for 
chemical systems under some assumptions. In particular, the Eigenfunction of the Hamiltonian will 
satisfy the NOCC equations. However, it is not possible to demonstrate that this solution is unique in 
general. In principle, Hamiltonians can be constructed that makes the $\bs{A}$-matrix in eq. 
\eqref{ddermat} singular, but this seems highly unlikely to occur in chemical systems. Interestingly, 
truncated NOCC does not satisfy the singles projection equations because the second term in eq. 
\eqref{firstterm} and the right hand side in eq. \eqref{thirdterm} are not zero. The same is true for 
OCC and it does not appear to have a large effect at least in well behaved systems\cite{NOCC,OCC}.

The advantage of NOCC compared to standard CC is that it is inherently gauge invariant, assuming a gauge 
invariant basis. While the effect of gauge dependence is usually small for most properties in CC theory, 
magnetic properties become origin dependent, resulting in unphysical behavior. This is very 
unsatisfying, especially when considering that CC is often the most accurate method available and used 
for benchmark calculations.

Truncated NOCC is unlikely to improve much on standard CC with respect to multireference 
system except for special cases. However, most multireference CC methods involve some sort of active 
space. This is especially true for single reference based multireference methods\cite{MultiRev}. 
Obtaining a good active space is critical in such methods and NOCC type orbital transformations makes 
it possible to include the transformation of the active space consistently in the Lagrangian. It would 
also be very interesting to see how the paired CC methods perform using NOCC orbitals instead of 
orthogonally optimized orbitals. 

Finally, it is worth noting that NOCC can be viewed as a special case of OACC where the entire orbital 
space is included in the active space\cite{OACC}. Orbital adapted CC makes it possible to obtain an 
optimal subspace of the Hilbert space spanned by the basis set. Basis set incompleteness is often the 
limiting factor for the accuracy of CC calculations and the OACC approach might make it possible to 
obtain greater accuracy at the same computational cost.
}

\begin{acknowledgments}

I would like to thank Simen Kvaal, Henrik Koch, Thomas Bondo Pedersen, Eirik Kj{\o}nstad and Fabian 
Faulstich for valuable discussions and input to this project. This work has received funding from 
ERC-STG-2014 under grant No 639508 and from the Research Council of Norway through its Centres of 
Excellence scheme, project number 262695.

\end{acknowledgments}

{\color{red}
\appendix
\section{Minimal example}\label{minex}

In this appendix, we will investigate the minimal example of two electrons in four spinorbitals. In 
chemistry, this corresponds to the hydrogen molecule in a minimal basis and we will assume a real 
symmetric Hamiltonian. Because the Hamiltonian does not couple singlet and triplet states, we do not 
have to worry about the triplet states if we assume our reference states are singlets because the 
cluster operators are reduced to a single singlet double excitation and its associated amplitude.
\begin{equation}\label{miniphi}
   \brat{\phi} = \bra{-}\ta_{1\al}\ta_{1\be} \quad
   \ket{\phi} = \ad_{1\be}\ad_{1\al}\ket{-} 
\end{equation}
\begin{equation}\label{miniTLa}
   \La = \la\ad_{1\al}\ad_{1\be}\ta_{2\be}\ta_{2\al} \quad
   T = \tau\ad_{2\al}\ad_{2\be}\ta_{1\be}\ta_{1\al} 
\end{equation}
Furthermore, products of the $T$-operator will be at least a quadruple excitation, so both the right 
and left wavefunctions will be linear in $T$, greatly simplifying the Lagrangian.
\begin{equation}\label{miniL}
\begin{aligned}
   L &= \brat{\Psi}H\ket{\Psi}\\ 
     &= \brat{\phi}(1+\La)\exp(-T)H\exp(T)\ket{\phi}\\
     &= \brat{\phi}(1+\La)(1-T)H(1+T)\ket{\phi}
\end{aligned}
\end{equation}
Taking the derivative of $L$ with respect to $\la$ gives us the equation for the amplitude $\tau$, 
resulting in a second order polynomial equation.
\begin{equation}\label{minilder}
\begin{aligned}
   \der{L}{\la} &= \brat{\mu_2}(1-T)H(1+T)\ket{\phi}\\
                &= \brat{\phi}H\ket{\phi} + \tau\bra{\ti{\mu}_2}H\ket{\mu_2}\\ 
                &- \tau\bra{\phi}H\ket{\phi} - \tau^2\bra{\phi}H\ket{\mu_2}\\
                &= \Htz + \tau(\Htt - \Hzz) - \tau^2\Hzt = 0\\
\end{aligned}
\end{equation}
\begin{equation}\label{minisqrt}
   \tau = \frac{(\Htt-\Hzz)\pm\sqrt{(\Htt - \Hzz)^2+4\Hzt\Htz}}{2\Hzt}
\end{equation}
If $H$ was symmetric, we would have $\Hzt=\Htz$ and the amplitude would be real given a real 
Hamiltonian. This is no longer guaranteed with a biorthogonal basis because $\Hzt$ and $\Htz$ can differ 
and, in principle, even have different signs. This seems unlikely to happen if we consider two cases of 
the hydrogen molecule; equilibrium bond length and the two atoms infinitely far apart. In the 
equilibrium case, the wavefunction is dominated by a single reference and the orbitals will be close to 
the canonical molecular orbitals. In such a case, $\Hzt\Htz$ will be positive and the two roots will 
correspond to the ground and excited state of the molecule. Because the terms under the square root are 
greater than the orbital difference, $\Htt-\Hzz$, the overall sign of the amplitude will depend on the 
choice of the root and $\Hzt$. Note that the magnitude of $\tau$ goes towards $0$ or $\infty$ as the 
difference in orbital energies increases, depending on the choice of root. In the infinitely stretched 
case, the orbital difference is zero and the expression is simplified. Again, the overall sign depends 
on the choice of root and $\Hzt$ and a symmetric matrix results in $\tau=1$.

Solving for $\la$ results in a linear equation that depends on $\tau$.
\begin{equation}\label{minitder}
\begin{aligned}
   \der{L}{\tau} &= \brat{\phi}(1+\La)(1-T)H\ket{\mu_2}\\
                 &- \la\brat{\phi}H(1+T)\ket{\phi}\\
                 &= \brat{\phi}H\ket{\mu_2} + \la\bra{\ti{\mu}_2}H\ket{\mu_2} 
                  - \la\tau\brat{\phi}H\ket{\mu_2}\\
                 &- \la\brat{\phi}H\ket{\phi} - \la\tau\brat{\phi}H\ket{\mu_2}\\
                 &= \Hzt + \la(\Htt-\Hzz) - 2\la\tau\Hzt = 0
\end{aligned}
\end{equation}
Inserting the expression for $\tau$ we again obtain an expression where the sign depends on the choice 
of root and $\Hzt$. Due to the intermediate normalization, the magnitude of $\la$ will always go to $0$ 
when the difference in orbital energies increases and will go to $\tfrac{1}{2}$ when the orbitals 
become degenerate and the Hamiltonian is symmetric.
\begin{equation}\label{minifrac}
\begin{aligned}
   \la &= \frac{\Hzt}{2\tau\Hzt - (\Htt-\Hzz)}\\
       &= \frac{\Hzt}{\pm\sqrt{(\Htt - \Hzz)^2+4\Hzt\Htz}}\\
\end{aligned}
\end{equation}
Importantly, the sign of $\la\tau$ will always be positive if $\Hzt\Htz$ is positive.
\begin{equation}\label{minilata}
   \la\tau = \frac{(\Htt-\Hzz)\pm\sqrt{(\Htt - \Hzz)^2+4\Hzt\Htz}}{\pm2\sqrt{(\Htt - \Hzz)^2+4\Hzt\Htz}}
\end{equation}
Finally, the $\bs{A}$ matrix from eq. \eqref{ddermat} becomes diagonal in this system. This can be 
realized by noting that there are no indices left to sum over in eqs. \eqref{example1} and 
\eqref{example2} for the off-diagonal elements. 
\begin{equation}\label{miniA}
   A_{ai,ai} = 1 + 2\la\tau
             = 2 \pm \frac{(\Htt-\Hzz)}{\sqrt{(\Htt - \Hzz)^2+4\Hzt\Htz}}
\end{equation}
Strictly speaking, there is no guarantee that $\Hzt\Htz$ is positive in the biorthogonal basis. However 
if the basis consists of two hydrogenic orbitals, the non-Hermitian terms in $\Hzt$ and $\Htz$ will be 
at least second order in $\bkappa$. Furthermore, in the case of infinitely stretched bond length, 
$A_{ai,ai} = 2$ unless $\Hzt\Htz$ also goes to zero somehow.
}

\end{document}